\documentclass[%
 aip,
 amsmath,amssymb,
 reprint,%
]{revtex4-1}

\usepackage{graphicx}
\usepackage{dcolumn}
\usepackage{bm}

\usepackage[utf8]{inputenc}
\usepackage[T1]{fontenc}
\usepackage{mathptmx}

\def\Tr#1{{\ensuremath{\text{Tr}{(#1)}}}}
\def\vec#1{{\ensuremath{{\bm{#1}}}}}
\def\n{\ensuremath{\bm{n} }}

\def\tensor#1{{\ensuremath{{\bm{#1}}}}}
\def\half{{\textstyle \frac{1}{2}}}

\def\fourth{{\textstyle \frac{1}{4}}}
\def\eighth{{\textstyle \frac{1}{8}}}
\def\threehalf{{\textstyle \frac{3}{2}}}
\def\twothirds{{\textstyle \frac{2}{3}}}
\def\threequarters{{\textstyle \frac{3}{4}}}
\def\threeeights{{\textstyle \frac{3}{8}}}
\def\p#1#2{{\ensuremath{\frac{\partial #1}{\partial #2}}}}
\usepackage{comment}

\begin{document}

\preprint{AIP/123-QED}

\title[]{Large deformation analysis of spontaneous twist and contraction in nematic elastomer fibres with helical director.}

\author{Andrea Giudici}
\email{ag2040@cam.ac.uk}
\author{John S. Biggins}%
\affiliation{ Department of Engineering, University of Cambridge, Trumpington St., Cambridge CB21PZ, U.K. }%

\date{\today}

\begin{abstract}
A cylindrical rubber fibre subject to twist will also elongate: a manifestation of Poynting's effect in large strain elasticity. Here, we construct an analogous treatment for an active rubber fibre actuated via an axisymmetric pattern of spontaneous distortion. We start by constructing an exact large-deformation solution to the  equations of elasticity for such fibre subject to imposed twist and stretch, which reveals spontaneous warping and twisting of the fibre cross-section absent in passive rubbers. We then compute the corresponding non-linear elastic energy, which encompasses the Poynting effect, but is minimized by a finite spontaneous twist and stretch. In the second half of the paper, we apply these results to understand the twist-contraction actuation of nematic elastomer fibres fabricated with director-fields that encode helical patterns of contraction on heating. We first consider patterns making a constant angle with respect to the local cylindrical coordinate system (conical spiral director curves) and verify the predicted spontaneous twist, contraction and cross-section deformation via finite elements. Secondly, we consider realistic director distributions for the experimentally reported fibres fabricated by cross-linking while simultaneously applying stretch and twist. Counter-intuitively, we find that maximum actuation twist is produced by applying a finite optimal twist during fabrication. Finally, we illustrate that spontaneously twisting fibres will coil into spring-like shapes on actuation if the ends are prevented from twisting relative to each other. Such twist-torsion coupling  would allow to make a tendril-like ``soft-spring'' actuator with low force and high linear stroke compared to the intrinsic contraction of the elastomer itself. 
\end{abstract}

\maketitle

\section{Introduction}

John Harrison invented the bi-metalic thermostat in 1759. Ever since, scientists and engineers have been deploying spatial patterns of spontaneous deformation to induce complex and dramatic actuation in solid materials. The thermal strains in metals and shape memory alloys are limited to a few percent, but, in recent years, the soft-matter community has demonstrated several systems in which patterns of geometrically large strains can be programmed into soft solids. Prominent examples include  patterns of swelling in gels \cite{klein2007shaping,kim2012designing,na2016grayscale, gladman2016biomimetic}, patterns of contraction in liquid crystal elastomers \cite{de2012engineering, ware2015voxelated, aharoni2018universal, barnes2019direct}, and patterns of inflation in ``baromorphs'' \cite{siefert2019bio, warner2020inflationary}. These large and exquisitely programmable shape changes can appear to bring the matter to life \cite{camacho2004fast, white2008high, gladman2016biomimetic}, and this is no coincidence, as they strongly resemble the patterns of muscular contraction that drive biological locomotion and the patterns of growth that underpin biological development \cite{thompson1942growth}.

This special issue focuses on the programming of such spontaneous shape changes in liquid crystal elastomers (LCEs). These are rubbery networks of rod-like mesogens which spontaneously align along a director, $\vec{n}$, to form a nematic phase \cite{warner2007liquid}. On heating or illumination, the nematic order can be disrupted (reflecting the nematic to isotropic transition in conventional liquid crystals, fig.\ \ref{fig:nematic}a) and, in LCEs, this transition is accompanied by a dramatic and reversible contraction  by a factor of $\lambda_s \sim 0.5$ parallel to the alignment director $\vec{n}$ \cite{warner2007liquid, kupfer1991nematic}.  LCEs are thus promising artificial muscles and soft actuators \cite{de1997artificial,wermter2001liquid}. 

Shape programming in LCEs is typically achieved by fabricating an elastomer in which the director $\n$ is spatially varying, generating a corresponding  pattern of contraction on heating. Director programming can be implemented by using surface-anchoring to pattern the director field in a nematic liquid sheet, and then crosslinking to form an elastomer \cite{de2012engineering, ware2015voxelated}.  Alternatively, one may use an aligning stress field to orient the director during crosslinking. This latter strategy was used to create the original globally aligned monodomain LCEs \cite{kupfer1991nematic}, and, much more recently, has been deployed to generate patterns of alignment during extrusion based 3-D printing \cite{ambulo2017four,kotikian20183d, lopez20184d} and during the direct shape programming of dual-network LCE sheets \cite{barnes2019direct}. 

The significant majority of work on LCE shape programming has focused on 2D sheets that bend and morph into curved surfaces on heating \cite{warner2020topographic,aharoni2018universal, mostajeran2016encoding,ware2015voxelated}. However, recently, \emph{Nocentini et al} demonstrated an LCE fibre that was twisted and stretched during cross-linking to imprint a helical director field\cite{Nocentini2017b}. On heating, the resulting helical contraction caused the fibre to spontaneously twist and contract. The design\cite{haines2016new,aziz2020torsional} and mechanics\cite{Charles2019} of other torsional artificial muscles has been the focus of much recent attention, as they offer a minaturizable version of a conventional rotary engine. However, these previous torsional muscles are fabricated by twisting multiple component fibres together\cite{foroughi2011torsional,Tondu2012,Zhang2012,Haines2014,Yuan2019} (for example by twisting passive in-extensible fibres around an inflating core \cite{Tondu2012,Zhang2012}) and their action relies on slip within the resultant fibre bundle. In contrast, the LCE torsional muscle is a monolithic cross-linked solid, and must be understood within the framework of misfit elasticity.

In this manuscript, we seek to construct such an elastic theory to predict and explain the spontaneous twist/stretch actuation of  \emph{Nocentini et al}'s LCE fibres. At first sight, the natural starting point is Timishenko's paradigmatic calculation of the curvature of a bimetalic strip\cite{timoshenko1925analysis}. Indeed, analogues of this small-strain analysis are frequently deployed to model LCE bilayers, and other spontaneously bending sheets and strips. 
Moreover, very recently, several authors have developed a corresponding theories of spontaneous bend and twist in elastic rods\cite{Aharoni2012,Kohn2018a,Moulton2020,Bauer2020,Cicalese2017}, and these certainly offer considerable insight into \emph{Nocentini et al's} LCE fibres.  However, all such theories are only valid for small spontaneous strains (or, more precisely, small incompatibilities of spontaneous strains), leading to the complete decoupling of stretch, twist and bend in the resultant elastic energies, and making such theories formally inapplicable to the large strains generated in LCEs. 

In contrast, in the field of rubber elasticity, there is a classic result, discovered by Poynting\cite{poynting1913changes} in 1913, that a rubber fibre that is twisted substantially will also stretch in response. This large strain effect cannot be captured by the small-strain approaches, but is captured by a simple and exact large-deformation solution for a twisted and stretched rubber cylinder \cite{Horgan2011,zurlo2020poynting}. Here, we derive an analogue of this exact large deformation solution for a rubber cylinder subject to an axisymetric pattern of spontaneous distortion. We find that these spontaneous distortions introduce simple modifications to the energy, so that it is minimized by an overall spontaneous twist and contraction. We then compute these spontaneous twists and contractions for various LCE fibres with different director fields, highlighting how a helical field with both azimuthal and longitudinal components is required to produce spontaneous twist. We validate our results with full 3-D finite element simulations. Finally, we estimate the spontaneous twist and stretch expected in fibres created by applying twist and stretch during cross-linking, as reported by \emph{Nocentini et. al.} \cite{Nocentini2017b}.  Counter-intuitively, we find that maximum spontaneous twist is achieved by an optimal finite  twist during cross-linking, with both too little or too much twist yielding lower performance. In conclusion, we discuss how these twisting fibres could be used to create coiling artificial muscles, with greatly amplified stroke compared to the intrinsic actuation of the LCE itself. 

\section{Twist and stretch of a rubber cylinder}\label{sec:simpletwist}
We start by recalling the classic large deformation solution for a twisted and stretched passive rubber cylinder\cite{Horgan2011,zurlo2020poynting}. More precisely, we consider a long cylindrical rubber fibre with undeformed radius $R_0$ and length $L \gg R_0$, that is subject to an angular twist $\Delta \theta$ between the two ends and an overall extension by a factor of $\lambda$. Working in cylindrical coordinates, if this deformation maps the material point  initially at $\vec{R}=(R,\Theta,Z)$ to the point $\vec{r}=(r,\theta,z)$, then the resultant deformation gradient is simply:
\begin{align}
\tensor{F}=\frac{\partial \vec{r}}{\partial \vec{R}}=\left(
\begin{array}{ccc}
 \frac{\partial r}{\partial R} & \frac{1}{R}\frac{\partial r}{ \partial \Theta}  & \frac{\partial r }{\partial Z}  \\
r \frac{\partial \theta}{\partial R} & \frac{r}{R}\frac{\partial \theta}{\partial \Theta} & r \frac{\partial \theta}{ \partial Z}  \\
 \frac{\partial z}{\partial R} &  \frac{1}{R}\frac{\partial z}{\partial \Theta} &  \frac{\partial z}{\partial Z}\\
\end{array}
\right).
\end{align}
In general, the hyper-elastic energy of a deformed solid may be written as
\begin{equation}
\mathcal{E}=\int_V W(\tensor{F})dV
\end{equation}
where $V$ is the volume in the reference configuration and $W$ is a energy density which depends on the deformation gradient. For an incompressible rubber, the simplest (neo-Hookean) energy density is given by 
\begin{equation}
W(\tensor{F})=\frac{1}{2}\mu \,\text{Tr}\left( \tensor{F}.\tensor{F}^T \right)+ p \left(\det(\tensor{F})-1\right),
\end{equation}
where $\mu$ is the shear modulus and the second term arises to impose volume conservation, $\det(\tensor{F})=1$, with $p$ being a spatially dependent Lagrange multiplier describing the pressure field in the material. 

Our challenge is to minimize this energy for the fibre subject to an overall imposed twist $\tau=\Delta \theta/L$ and longitudinal stretch $\lambda$. Since we are looking for states of uniform stretch and twist along the length of the cylinder, the outer surface must have the form
\begin{equation}
z(R_0,\Theta,Z)=\lambda Z\,\,\,\,\,\,,\,\,\,\,\,\,\,\theta(R_0,\Theta,Z)=\Theta+\tau Z.
\label{BC}
\end{equation}
These forms clarify that $\tau=\frac{d\theta}{dZ}$ corresponds to the angular twist per unit \emph{reference state} length. 

\begin{figure}[h]
\begin{center}
\includegraphics[width=8 cm]{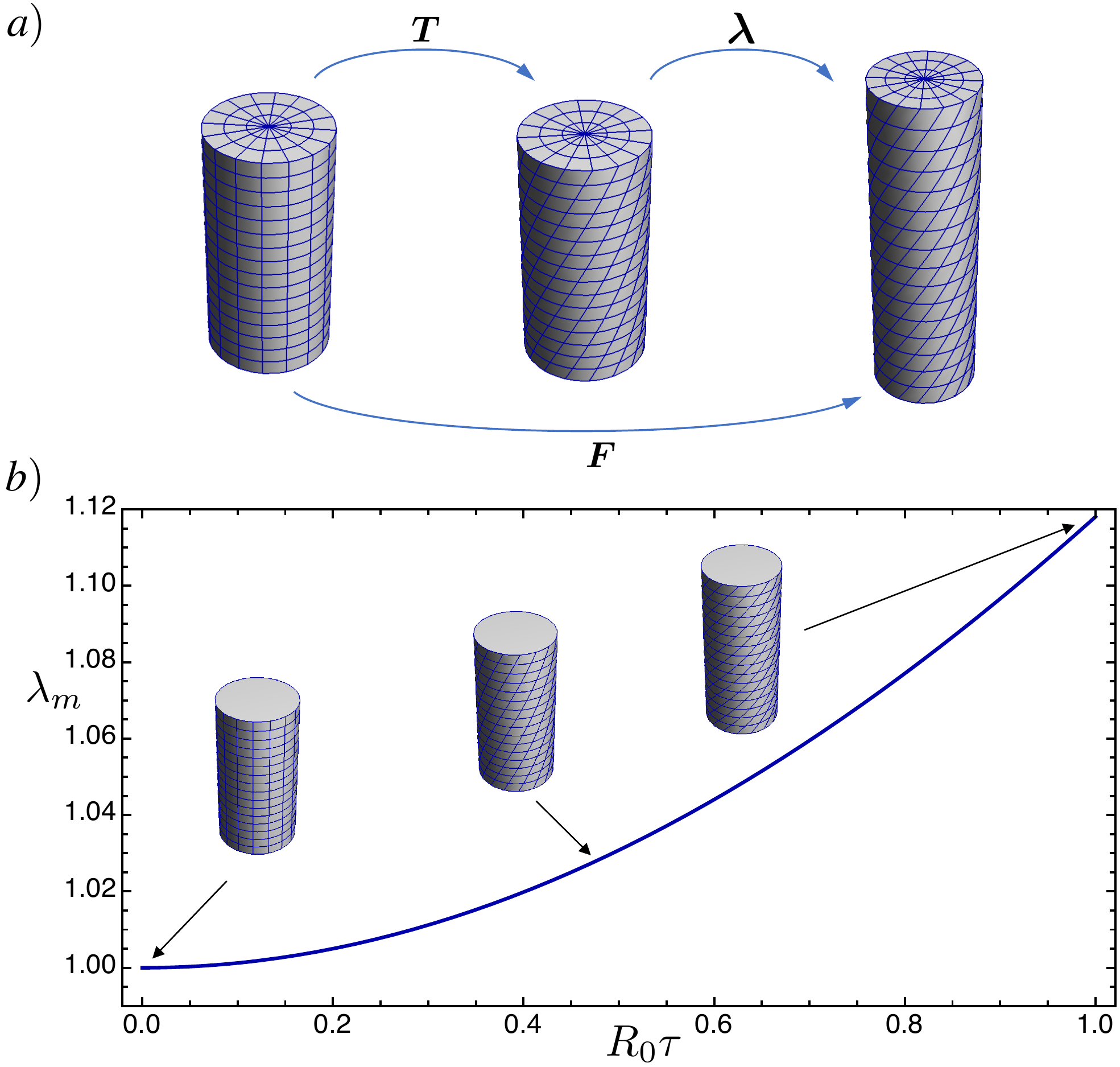}
\end{center}
\caption{\small a) Schematics of the decomposition of $\tensor{F}=\tensor{T}\cdot \tensor{\lambda}$. b) Poynting effect: the equilibrium stretch increases as more twist is imposed on the system.}
\label{poyinting}
\end{figure}

Minimising the elastic energy with respect to variations in $\vec{r}(\vec{R})$  leads to the traditional bulk equations of mechanical equilibrium
\begin{equation}
\nabla \cdot \tensor{\Sigma}=0,
\label{eq1}
\end{equation}
and free boundary condition 
\begin{equation}
\tensor{\Sigma} \cdot \hat{\vec{R}}=0
\label{eq2}
\end{equation}
where the first Piola-Kirchhoff stress tensor is given by
\begin{equation}
\tensor{\Sigma} \equiv \p{W}{\tensor{F}}=\mu \tensor{F}+p \det(\tensor{F})\tensor{F}^{-T},
\label{pk1}
\end{equation}
Finally, minimising with respect to variations in $p$ returns the expected bulk condition of impressibility
\begin{equation}
\det(\tensor{F})=1.
\label{eq3}
\end{equation}

In the case of the twisted stretched fibre, the simplest possible fields are those with homogeneous twist and (isochoric) stretch
\begin{align}
\notag
r(R,\Theta,Z)&=R/\sqrt{\lambda}\\
\notag
\theta(R,\Theta,Z)&=\Theta+\tau Z\\
\notag
z(R,\Theta,Z)&=\lambda  Z\\
\notag
p(R,\Theta,Z)&=-1/\lambda,\\
\label{simpletwist}
\end{align}
which do indeed solve all the relevant bulk and boundary conditions.  The associated deformation gradient is simply
\begin{align}
\tensor{F}=\left(
\begin{array}{ccc}
\lambda^{-1/2} & 0  & 0  \\
0 &  \lambda^{-1/2}  & R \tau\lambda^{-1/2}  \\
0 &  0 &  \lambda\\
\end{array}
\right),
\label{Fgrad}
\end{align}
and, upon substituting this back into the elastic energy, we obtain the total energy of a twisted and stretched fibre as:
\begin{equation}
\frac{\mathcal{E}}{\mu\pi L R_0^2}=\frac{1}{2}\left( \lambda ^2 +\frac{2}{\lambda}+\frac{R_0^2 \tau ^2}{2\lambda}\right).
\label{energy0}
\end{equation}
We note that the first two terms in the parenthesis are simply the familiar uni-axial stretching energy for a neo-Hookean rubber, while the final term determines the energy cost of twisting the fibre. The energy is trivially minimised by $\tau=0$ for all values of $\lambda$, but the minimum stretch is given by
\begin{equation}
\lambda_{m}=\sqrt[3]{\frac{(R_0 \tau)^2+4}{4}}\geq1,
\end{equation}
with $\lambda_{m}=1$ only when $\tau=0$, as illustrated in fig.\ \ref{poyinting}b. This asymmetric coupling between the two quantities is a direct manifestation of the Poynting effect \cite{Billington1986} in nonlinear elasticity. To clarify its origin, we note that the deformation gradient can be achieved as a pure twist followed by a pure stretch, as illustrated in figure  \ref{poyinting}a. Mathematically, this corresponds to decomposing the deformation gradient as
 \begin{equation}
 \tensor{F}=\tensor{\lambda}.\tensor{T},
 \label{decomp1}
 \end{equation}
 where
 \begin{equation}
 \tensor{\lambda}=\text{diag}(\lambda^{-1/2},\lambda^{-1/2},\lambda)\,\,\,\,\,\, \text{,}\,\,\,\,\, \tensor{T}=\tensor{\delta}+R\tau  \,\,\hat{\vec{e}}_r \hat{\vec{e}}_{\Theta},
 \label{tensors}
 \end{equation}
 and $\tensor{\delta}$ is the identity matrix. From the form of $\tensor{T}$, we see that a smaller radius implies a smaller deformation for the same twist. Indeed, the twist energy $R_0 \tau^2/\lambda$ can  be written as $R_f^2 \tau^2$ where $R_f=R_0/\sqrt{\lambda}$ is the final radius of the fibre. These consideration suggest that twist energy is partially relieved by stretch as this reduces the radius via Poisson effects.  The resultant twist-stretch coupling is a paradigmatic example of the inherent geometric non-linearity of large strains.

\section{Twist and stretch induced by a spontaneous deformation.}
\subsection{Spontaneous deformation field}
In light of what we learned from the simple twisting case, we now turn our attention to a cylindrical rubber fibre that undergoes a  heterogeneous spontaneous distortion, such that, locally, the energy minimizing deformation is given by $\tensor{F}=\tensor{G}(R,\Theta,Z)$. Since our ultimate motivation is to understand the spontaneous twist and stretch of nematic LCE fibres, we restrict consideration to $\tensor{G}$ that are axisymetric, isochoric, and independent of $Z$. However, the pattern of spontaneous distortion is allowed to be incompatible, so that the cylinder cannot attain $\tensor{F}=\tensor{G}$ throughout, but will instead relax to an internally stressed state that minimizes the total elastic energy. 

If the actual local deformation from the original state, prior to spontaneous distortion is $\tensor{F}$, then the elastic deformation from the local relaxed state is simply $\tensor{F}\cdot\tensor{G}^{-1}$, where the second term reverses the effect of spontaneous deformation, and the first applies the actual deformation. The new elastic energy of the fibre, after spontaneous distortions, is thus
\begin{equation}
\mathcal{E}=\int_V\left[\half \mu \Tr{\tensor{F}\cdot\tensor{G}^{-1}\cdot \tensor{G}^{-T}\cdot\tensor{F}^T} +p \,(\det(\tensor{F})-1) \right]dV.
\label{enG}
\end{equation}
 This "multiplicative decomposition'' form\cite{dicarlo2002growth} was first introduced for elasto-plastic deformations\cite{lee1967finite} and now pervades and unifies the study of solids with spontaneous deformations, including growing tissues \cite{dervaux2008morphogenesis, tallinen2014gyrification}, swelling gels \cite{dervaux2011buckling, tallinen2013surface}, thermal expansion \cite{vujovsevic2002finite, Kohn2018a} and, as we shall clarify later, nematic elastomers heated to the isotropic state.
Importantly, the spontaneous deformation only affects the energy via the  combination  $\tensor{g}=\tensor{G}^{-1}\cdot \tensor{G}^{-T}$ (corresponding to the Finger tensor of $\tensor{G}$) which, in our fibers, takes the symmetric and axisymetric form
\begin{equation}
\tensor{g}=\left(
\begin{array}{ccc}
g_{R R}(R) & g_{R \Theta}(R)  & g_{R Z}(R)   \\
g_{R \Theta}(R)    & g_{\Theta\Theta}(R) & g_{\Theta Z}(R)   \\
g_{R Z}(R) &  g_{\Theta Z}(R)&g_{Z Z}(R)\\
\end{array}
\right).
\end{equation}
The isochoric condition on all rubbers, including LCEs, requires $\det{(\tensor{g})}=\det(\tensor{G})=1$.

\subsection{Resultant deformation fields}
Minimising \eqref{enG} with respect to variations in $\vec{r}(\vec{R})$ and $p(\vec{R})$ leads to the same bulk and boundary eqns as before  (\eqref{eq1}, \eqref{eq2}, \eqref{eq3}), but with the new  Piola-Kirchhoff tensor given by
\begin{equation}
\tensor{\Sigma}=\mu \tensor{F}\cdot \tensor{g}+p \det(\tensor{F}) \tensor{F}^{-T}.
\end{equation}
To solve for the actual deformation field, we first observe that, given axisymmetry, $Z$ independence and incompresibility, the deformation must take the simple form  
\begin{align}
\notag
r(R,\Theta,Z)&=R/\sqrt{\lambda}\\
\notag
\theta(R,\Theta,Z)&=\Theta+\tau \left(Z+f_{\theta}(R)\right)\\
\notag
z(R,\Theta,Z)&=\lambda  (Z+f_{z}(R))\\
p(R,\Theta,Z)&=-f_p(R)/\lambda.
\label{fields1}
\end{align}
where $f_\theta$, $f_z$ and $f_p$ are as-yet unknown functions of $R$, which modify the solution from the original one for a passive rubber cylinder. The $f_{\theta}$ term introduces rotation as a function of radius, and allows radii in the reference configuration to become curves in the final configuration. Similarly, the $f_{z}$ term  encodes warping of each cross section into an identical surface of revolution, while $f_p$ allows the pressure to vary with radius. 

\begin{figure}[h]
\begin{center}
\includegraphics[width=8 cm]{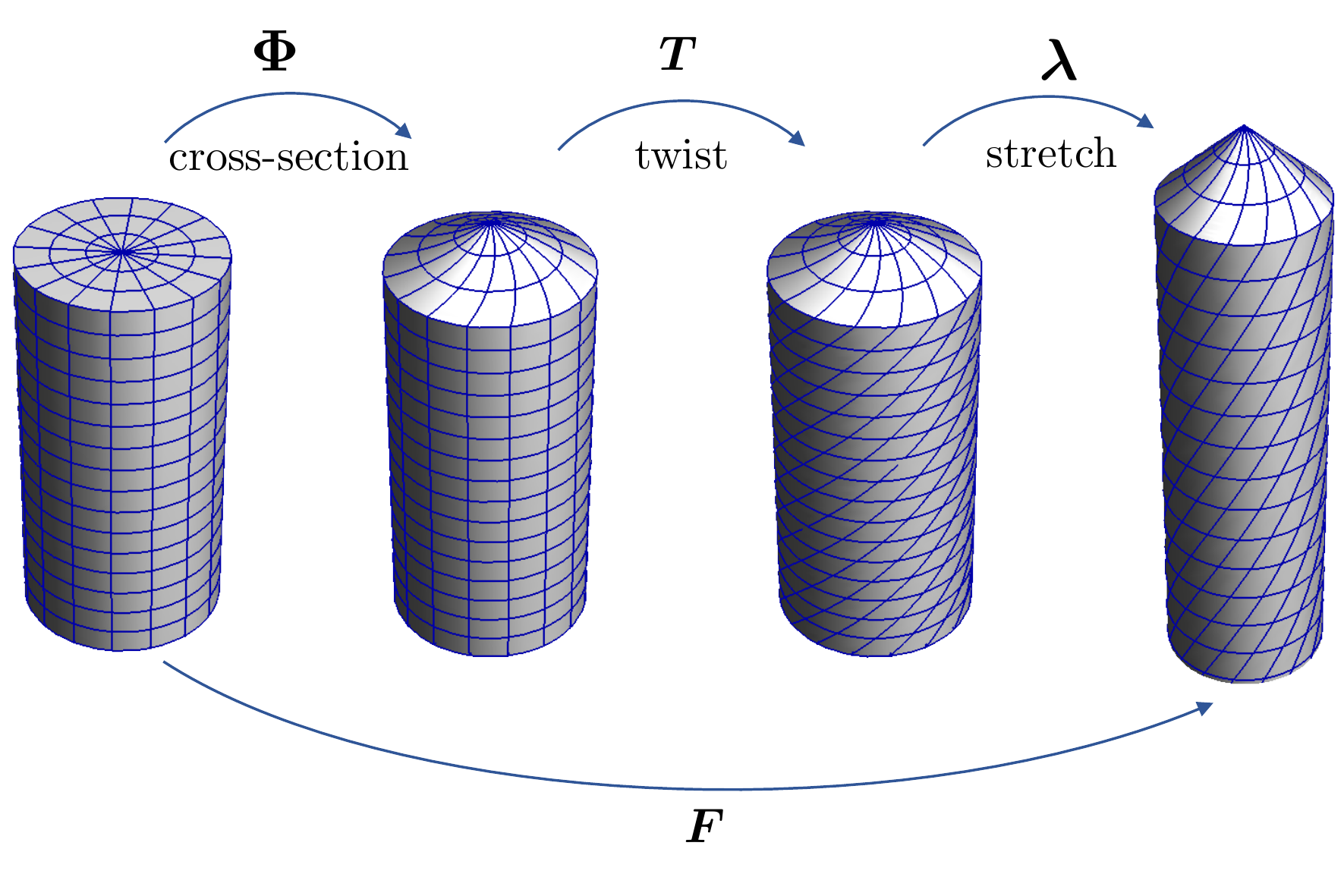}
\end{center}
\caption{\small a) Schematics of the decomposition of $\tensor{F}=\tensor{T}\cdot \tensor{\lambda} \cdot \tensor{\Phi}$. }
\label{fig:decomposition}
\end{figure}

At this stage, one could methodically substitute these fields into the bulk and boundary equations, and then solve for $f_\theta$, $f_z$ and $f_p$. However,  one can substantially simplify this process by noting that the resultant deformation field can now be decomposed as
\begin{align}
\tensor{F}& =\left(
\begin{array}{ccc}
 \frac{1}{\sqrt{\lambda }} & 0 & 0 \\
 \frac{R f_{\theta }'}{\sqrt{\lambda }} & \frac{1}{\sqrt{\lambda }} & \frac{R \tau }{\sqrt{\lambda }} \\
 \lambda  f_z' & 0 & \lambda  \\
\end{array}
\right)=\tensor{\lambda} \cdot \tensor{T} \cdot \tensor{\Phi},
\label{Fdecomp}
\end{align}
where $\tensor{\lambda}$ and $\tensor{T}$ are again pure stretch and twist (eq.\ \eqref{tensors}) while
\begin{equation}
\tensor{\Phi}=\left(
\begin{array}{ccc}
 1 & 0 & 0 \\
 R \tau \left( f_{\theta }'- f_z'\right) & 1 & 0 \\
 f_z' & 0 & 1 \\
\end{array}
\right)
\end{equation}
describes the deformation of the cross-section in the absence of twist or stretch, and is the only part that depends on $f_\theta$ and $f_z$. This decomposition is illustrated in Fig.\ \ref{fig:decomposition}

Furthermore, since only derivatives of the two fields $f_\theta$ and $f_z$ appear in $\tensor{F}$, we can substitute this deformation into the energy, and minimise directly with respect to variations in $f'_z(R)$ and $f'_{\theta}(R)$, leading to the conditions:
\begin{align}
\p{W}{f'_{\theta}}&=\mu \mathrm{Tr} \left(\tensor{\lambda} \cdot \tensor{T} \cdot \tensor{\Phi}\cdot \tensor{g}\cdot \p{\tensor{\Phi}^T}{f'_{\theta}}\cdot \tensor{T}^T\cdot \tensor{\lambda}^T\right)=0\\
\p{W}{f'_{z}}&=\mu \mathrm{Tr} \left(\tensor{\lambda} \cdot \tensor{T} \cdot \tensor{\Phi}\cdot \tensor{g}\cdot \p{\tensor{\Phi}^T}{f'_{z}}\cdot \tensor{T}^T\cdot \tensor{\lambda}^T\right)=0.
\end{align}
Evaluating these yields the simple uncoupled differential equations
\begin{align}
g_{R \Theta}+R \tau g_{R R} f'_{\theta}+R \tau g_{R Z}&=0,\\
\lambda ^2\left(g_{R R} f'_z+g_{R Z}\right)&=0,
\end{align}
which can be directly integrated to obtain the fields
\begin{align}
\notag
f_{\theta}&= -\int _{R_0}^{R}\frac{g_{R Z}(u)}{g_{R R}(u)} du -\tau^{-1}\int _{R_0}^{R}\frac{1}{u}\left(\frac{g_{R\Theta }(u)}{\,g_{R R}(u)}\right)du,\\
\notag
f_z&=- \int _{R_0}^{R}\frac{g_{R Z}(u)}{g_{R R}(u)} du.
\end{align}
These solutions then imply the full form of the deformation fields:
\begin{align}
\notag
r&=R/\sqrt{\lambda}\\
\notag
\theta&= \tau\,z/\lambda-\int _{R_0}^{R}\frac{1}{u}\left(\frac{g_{r\Theta }(u)}{\,g_{R R}(u)}\right)du,\\
z&=\lambda \left(Z- \int _{R_0}^{R}\frac{g_{R Z}(u)}{g_{R R}(u)} \right)du.
\end{align}
Finally, if we now substitute these back into the original bulk and boundary equations (\eqref{eq1}  and \eqref{eq2}) we can confirm they are fully solved (e.g.\ in mathematica) provided the pressure is taken as 
\begin{align}
f_p(R)&=\tau \int_{{R0}}^R f_{\theta }'(u) \left(\tau \,u f_{\theta }'(u) g_{{RR}}(u)+2 \tau  u g_{{RZ}}(u)+2 g_{R\Theta
   }(u)\right) \, du\notag \\
   \notag
   &\,\,\,\,\,-g_{{RR}}(R)+\int_{{R0}}^R \left(g_{\Theta \Theta }(u)-g_{R R}(u) \right)u^{-1} \, du\\
   &\,\,\,\,\,+\int_{{R0}}^R \tau  \left(2
   g_{{\Theta Z}}(u)+\tau  u g_{{ZZ}}(u)\right) \, du.
\end{align}

\subsection{Twist and stretch elastic energy}
We now substitute the fields into the elastic energy (eqn.\ \ref{enG}) to evaluate the energy for a rubber rod with imposed twist $\tau$, stretch $\lambda$ and a pattern of spontaneous deformation $\tensor{g}$. To do this, it is convenient to use the decomposition in \eqref{Fdecomp}. We then see immediately that the effects of $\tensor{g}$ in the energy are entirely contained within the symmetric tensor
\begin{align}
\tilde{\tensor{g}}=\tensor{\Phi} \cdot \tensor{g} \cdot \tensor{\Phi}^T=\left(
\begin{array}{ccc}
 g_{R R} & 0 & 0 \\
 0 & g_{\Theta \Theta }-\frac{g_{R\Theta }^2}{g_{{RR}}} &
   g_{{\theta Z}}-\frac{g_{{R\Theta }} g_{{RZ}}}{g_{{RR}}}
   \\
 0 & g_{{\Theta Z}}-\frac{g_{{R\Theta}}
   g_{{RZ}}}{g_{{RR}}} &
   g_{{ZZ}}-\frac{g_{{RZ}}^2}{g_{{RR}}} \\
\end{array}
\right),
\end{align}
which captures the residual part of $\tensor{g}$ after allowing the cross-sections to relax via $\tensor{\Phi}$, but in the absence of twist or stretch. By inspection we may then multiply out all the terms in the energy to find:
\begin{align}
\frac{\mathcal{E}}{\mu \pi L R_0^2}&=\frac{a_0 }{\lambda} +a_1 \lambda^2 +\frac{b}{\lambda} \tau +\frac{c}{\lambda} \tau^2
\end{align}

with the coefficients given by:
\begin{align}
\notag
a_0&=\half \int_0^{R_0} \left(\tilde{\tensor{g}}_{RR}+\tilde{\tensor{g}}_{\Theta \Theta}\right) R dR\\
\notag
a_1&=\half \int_0^{R_0} \tilde{\tensor{g}}_{ZZ} R dR\\
\notag
b&=\int_0^{R_0} \tilde{\tensor{g}}_{\Theta Z}R^2 dR\\
c&=\half \int_0^{R_0} \tilde{\tensor{g}}_{ZZ}R^3 dR.
\label{coeff1}
\end{align}
 We note that these terms are related to the zeroth, first and second moments of aspects of the spontaneous distortion, reminiscent of those derived via Gamma convergence for linear elastic rods \cite{Kohn2018a}.
 
As in the simple case explored in section \ref{sec:simpletwist}, twist and stretch are coupled via the $\lambda^{-1}$ in the twist energy. However, with the addition of the linear term in $\tau$, when $b \neq 0$ the energy is minimised by a non-zero twist $\tau_m$ and a finite stretch $\lambda_m$ given by:
\begin{equation}
\tau_{m}=-\frac{b}{2 c}\,,\,\,\,\,\,\,\,\,\,\,\lambda_{m}=\sqrt[3]{\frac{4 a_0 c-b^2}{2 a_1 c}}.
\label{minimisers}
\end{equation} 
We can use these and $\bar{\lambda}=\lambda/\lambda_m$ to rewrite an energy with the same structure as in eqn \eqref{energy0}, but now including the source terms,
\begin{align}
\frac{\mathcal{E}}{\mu \pi L R_0^2}&= \lambda_m^2\left[a_1\left(\frac{2}{\bar{\lambda}}+\bar{\lambda}^2 \right)+\frac{(c/\lambda_m^3)}{\bar{\lambda}}(\tau-\tau_m)^2\right].
\label{eqn:energymain}
\end{align}
This is our main result and captures the emergent twist/stretch nature of  fibres with a spontaneous deformation field. 

Thus, our calculation reveals that an axisymmetric spontaneous deformation can induce 4 key effects in a fibre:\\
    1. spontaneous length change.\\
    2. spontaneous twist.\\
    3. cross-section warping, via the function $f_z$, so initial disks become surfaces of revolution.\\
    4. cross-section twisting via $f_{\theta}$, so initial radii become curves.\\
These effects stem from different aspects of the spontaneous deformation, and may or may not co-occur in a given case.

In the following sections, we concentrate on spontaneous deformation fields arising from a LCE undergoing nematic-isotropic transition. 

\section{Nematic elastomer fibres}

\begin{figure}[b]
\begin{center}
\includegraphics[width=8 cm]{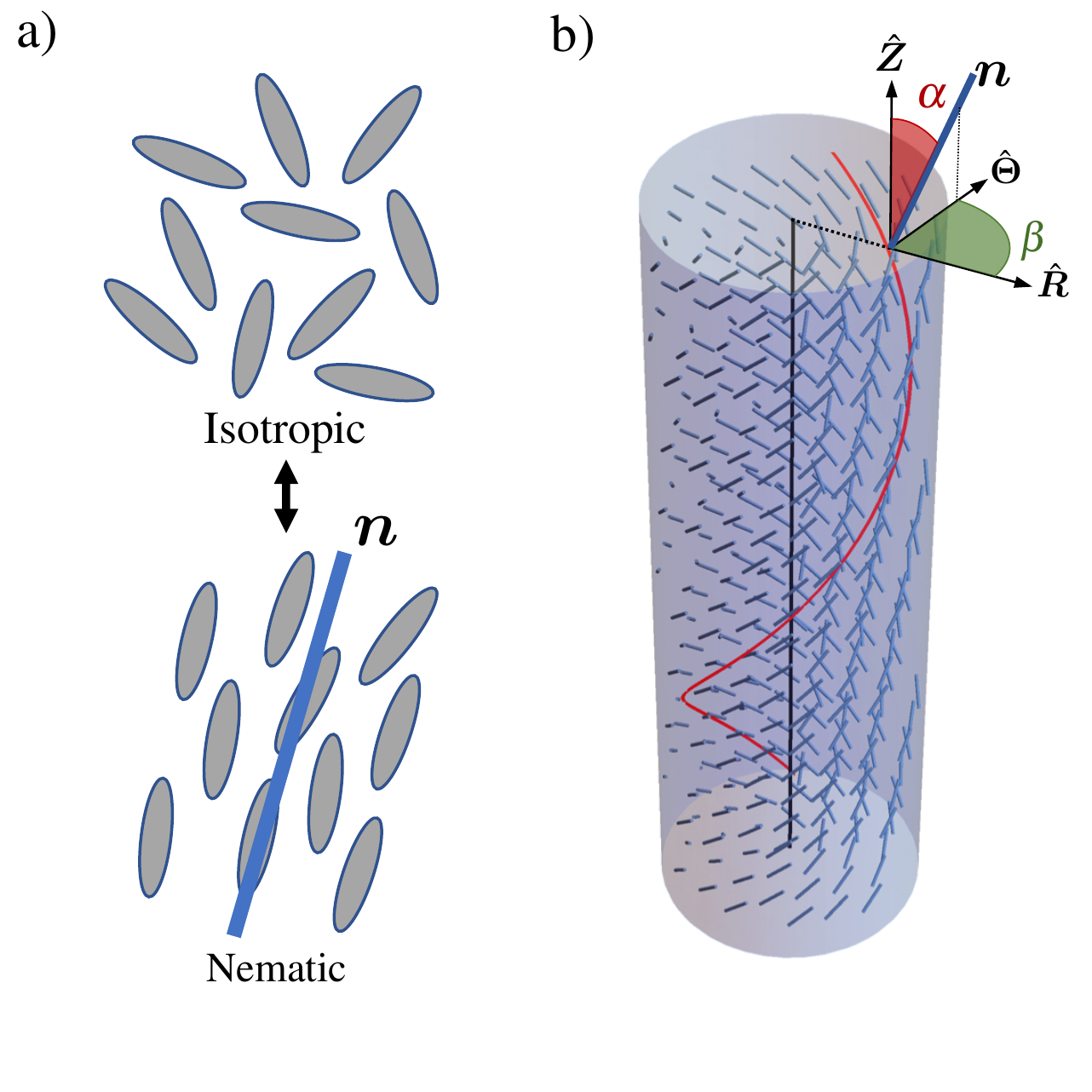}
\end{center}
\caption{\small a) Schematics of isotropic-nematic transition and the resulting alignment along $\vec{n}$. b) Example of a $R$-independent director field with $\alpha=\pi/4$ and $\beta=\pi/2$, as shown on the top face. In red, we mark a helical integral curve to highlight the chirality of the director field.}
\label{fig:nematic}
\end{figure}

\begin{figure*}
\includegraphics[width=\textwidth]{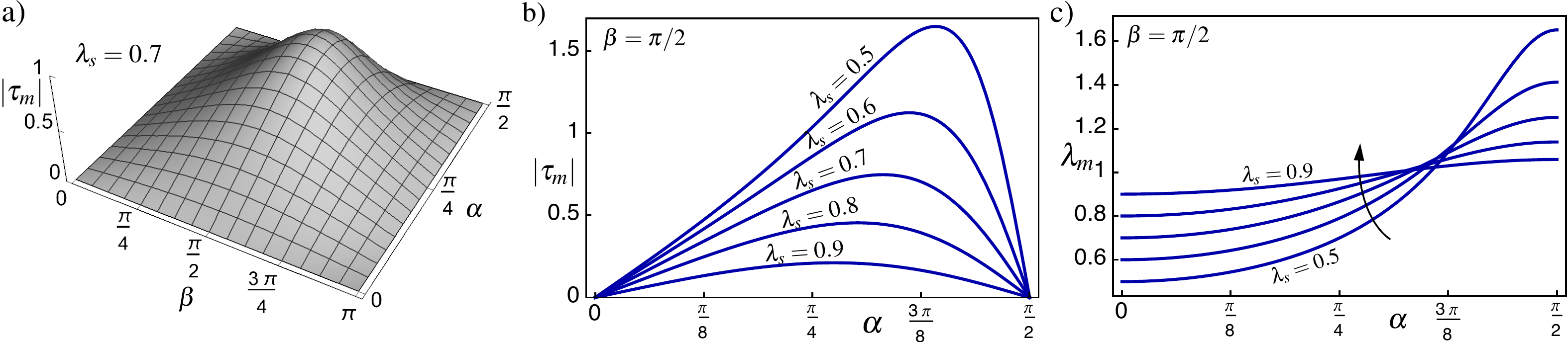}
\caption{\label{fig:costant} a) 2D plot of twist as a function of both angles $\alpha$ and $\beta$. The twist is maximised when $\beta=\pi/2$ and $\alpha \sim \pi/4$. b) -- c) Variation of the twist and stretch as a function of $\alpha$ for different values of $\lambda_s$.}
\end{figure*}
We want to apply our theoretical machinery to nematic elastomer cylinders, encoded with a spatially variable director pattern $\vec{n}$. On heating through the nematic-isotropic transition schematically shown in fig \ref{fig:nematic} a), the elastomer will spontaneously contract by a factor $\lambda_s$ along the director and, to preserve volume, elongate by $1/\lambda_s$ in the two perpendicular directions, corresponding to a spontaneous deformation  $\tensor{G}=\text{diag}(\lambda_s^{-1/2},\lambda_s^{-1/2},\lambda_s)$ in a frame aligned with the director \cite{warner2007liquid}. We can thus write $\tensor{g}$ for such spontaneous actuation as:
\begin{equation}
\tensor{g}=\lambda_s^{-2} \vec{n}\vec{n}+ \lambda_s(\tensor{\delta}-\vec{n}\vec{n}).
\label{nematicG2}
\end{equation}
As discussed in Appendix \ref{appendixa}, the resulting pre-strained neo-Hookean exactly reproduces the ``trace-formula'' energy commonly encountered in the LCE literature\cite{warner2007liquid} for an LCE cross-linked in the nematic state then heated to the isotropic state. This form neglects stress induced changes in the degree of nematic alignment, which is appropriate in the isotropic phase, where an LCE is simply a classical rubber provided one is not too close to the transition temperature. \emph{Nocentini et al}'s  experimental LCE undergoes its transition in a 10K window around 373K, and is heated to 393K for actuation, allowing for this approximation. We also note that the nematic actuation strains during the transition, $\delta L/L \sim 0.3$, vastly exceed other thermal effects such as (linear) thermal expansion which has characteristic size $\delta L/L \sim 10^{-5}K^{-1}$.




We again limit attention to axisymmetric director patterns. In cylindrical coordinates, the (unit) director field is then described by two angles  $\alpha(R)$ and $\beta(R)$ (fig \ref{fig:nematic} b) so that
\begin{equation}
    \vec{n}=\cos \beta \sin \alpha \vec{\hat{e}}_R+\sin \beta \sin \alpha \vec{\hat{e}}_\Theta+\cos \alpha \vec{\hat{e}}_Z.
\end{equation}
In what follows, we first consider the case of $\alpha$ and $\beta$ independent of $R$, and show that spontaneous twist requires an oblique director field between azimuthal and longitudinal directions, while warping of the cross-sections requires a director that is oblique between the longitudinal and radial directions. Finally, we  consider some more realistic $R$ dependent director fields for the fibres created by twisting and stretching during cross-linking\cite{Nocentini2017b}, and show that spontaneous twisting is maximized by a finite optimum degree of twisting at genesis.

\subsection{Example: R-independent Director field.}

\subsubsection{Theoretical predictions}
We start by considering director patterns in which the angles $\alpha$ and $\beta$ are constants, independent of $R$. In this case, the integrals in our solutions can be conducted analytically and yield the displacement fields
\begin{align}
\notag
r&=R/\sqrt{\lambda}\\
\notag
\theta&=\Theta+\tau \left(Z +\frac{g_{R Z}}{g_{R R}}(R_0-R)\right)-\log(R/R_0)\frac{g_{R \Theta}}{g_{R R}}\\
\notag
z&=\lambda\left(Z +\frac{g_{R Z}}{g_{R R}}(R_0-R)\right).
\end{align}
From the linear dependence on $R$ in $z$ we note that flat disc cross sections turn into cones, while $\theta(R)$ shows that radii turn into spirals which tend to conical-spirals at the centre of the fibre. Though such singular structures may seem surprising at first glance, a more careful look reveals that integral curves of the director-field  $\vec{n}$ are also conical spirals making a constant angle $\beta$ with the radial direction. Furthermore, in planar LCEs encoded with constant angle +1 defects, the integral curves also form planar log spirals (the planar projection of a conical spiral) and it is well established that the resultant actuation transforms the sheet into twisted conical surfaces in which the radii transform into conical spirals\cite{modescones,mostajeran2016encoding, warner2018nematic}.

In the simple case in which $\beta=\pi/2$, the integral curves become simple helices, as shown in Fig.\ \ref{fig:nematic} b), and are helpful to understand how twist is developed during activation. In the most simple sketch of the mechanics, heat/illumination drives a contraction by $\lambda_s$ along the integral curve as well as an increase in its radius by a factor of $\sqrt{\lambda_s}$. The twisting thus occurs to reduce the length of the integral curve, with the integral curve playing the same role of a sub-fibre in twisted fibre bundle torsional muscles\cite{foroughi2011torsional,Tondu2012,Zhang2012,Haines2014,Yuan2019}.  

Given constant $\alpha$ and $\beta$, the coefficients in our energy (eqn. \eqref{eqn:energymain}) are:
\begin{align}
\notag
a_0=&\eighth \lambda _s^{-2}  R_0^2d2  \sin ^4\alpha  \sin ^2 2 \beta\left(1\!-\!\lambda _s^3\right) \\
\notag
&\,\,\,\,\,\,+\eighth \lambda _s^{-2}  R_0^2\left(\!3 \lambda _s^3\!+\!1\!+\cos 2 \alpha (\lambda_s^3-1)\right)\\
   \notag
  a_1=&\fourth R_0^2 \lambda_s d\left(3\!+\!\lambda_s\!-\! \left(\cos 2\alpha \!+\!2 \cos 2\beta \sin^2 \alpha  \right)(\lambda_s^3\!-\!1) \right)\\
   \notag
   b=&\twothirds R_0^3 d \sin 2 \alpha  \sin
   \beta  \lambda _s \left(1-\lambda _s^3\right)\\
   c=&\eighth R_0^4 \lambda _s d \left(3\!+\lambda_s\!+\!\left(\cos 2\alpha\!+\!2 \cos 2 \beta \sin^2 \alpha  \right)\left(1-\lambda _s^3\right) \right)
   \label{eq:coeff2}
\end{align}
where
\begin{align}
\notag
d=\left(2 \sin ^2\alpha  \cos 2 \beta 
   \left(1-\lambda _s^3\right)+\cos 2 \alpha  \left(\lambda _s^3-1\right)+3
   \lambda _s^3+1\right)^{-1}.
\end{align}
These can be used together with eqns.  \eqref{nematicG2} and \eqref{minimisers},  to obtain the spontaneous twist:
\begin{equation}
\tau_m=\frac{-8  \sin 2 \alpha \, \sin \beta }{R_0 \left( 6 \sin ^2\alpha  \cos 2 \beta \!+\! 3 \cos 2 \alpha \!-\! 3 \left(\lambda_s^3 \!+\! 3\right)/\left(\lambda_s^3\!-\!1\right)\right)}.
\end{equation}
The behaviour of the twist as a function of $\alpha$ and $\beta$ is shown in Figure \ref{fig:costant} a) and b) for typical values during a nematic-isotropic (heating) transition ($1>\lambda_s>0.5$). Importantly, we see that the twist vanishes when $\beta=0$ or $\alpha=0,\,\pi/2$, corresponding to a purely longitudinal or azimuthathal-radial director-field where no shear $\Theta$-$Z$ is present. On the other hand, $\tau_m$ is maximised when $\beta=\pi/2$ (independent of $\alpha$ and $\lambda_s$), indicating a director with no radial component. The largest possible twist is then given by choosing $\alpha=\frac{1}{2} \cos ^{-1}\left(\frac{\lambda_s^3-1}{\lambda_s^3+1}\right)\sim \pi/4+\threequarters (\lambda_s-1)+O((\lambda_s-1)^2)$, showing that maximum twist is achieved at an oblique angle biased towards the azimuthal direction for larger $\lambda_s$, as shown in Fig \ref{fig:costant} b).

Similarly, the overall spontaneous stretch, $\lambda_m$, is given by eqn\ \eqref{minimisers}. Although the full expression is too complicated to reproduce here, in the simple and twist-maximising case of $\beta=\pi/2$, it reduces to 
\begin{equation}
\notag
\lambda_m\!=\!\frac{36\! \cos\! 2 \alpha  \left(\!1\!-\!\lambda _s^3\!\right) \lambda _s^3\!-\!\cos \!4 \alpha  \left(\!\lambda _s^3\!-\!1\!\right)^2\!+\!37 \lambda _s^6\!+\!106 \lambda
   _s^3\!+\!\!1}{36 \left(\cos 2 \alpha  \left(1-\lambda _s^3\right)+\lambda _s^3+1\right)^{2}}.
\end{equation}
In Fig. \ref{fig:costant} c), we can see how this stretch behaves between the two extremes of a longitudinally aligned ($\alpha=0$) director field, yielding a simple contraction of $\lambda_m=\lambda_s$, and of a azimuthal director field yielding $\lambda_m=\left(\half \left(1+\lambda_s \right)/\lambda_s^3\right)^{1/3}$.
\begin{figure}[b]\includegraphics[width=\columnwidth]{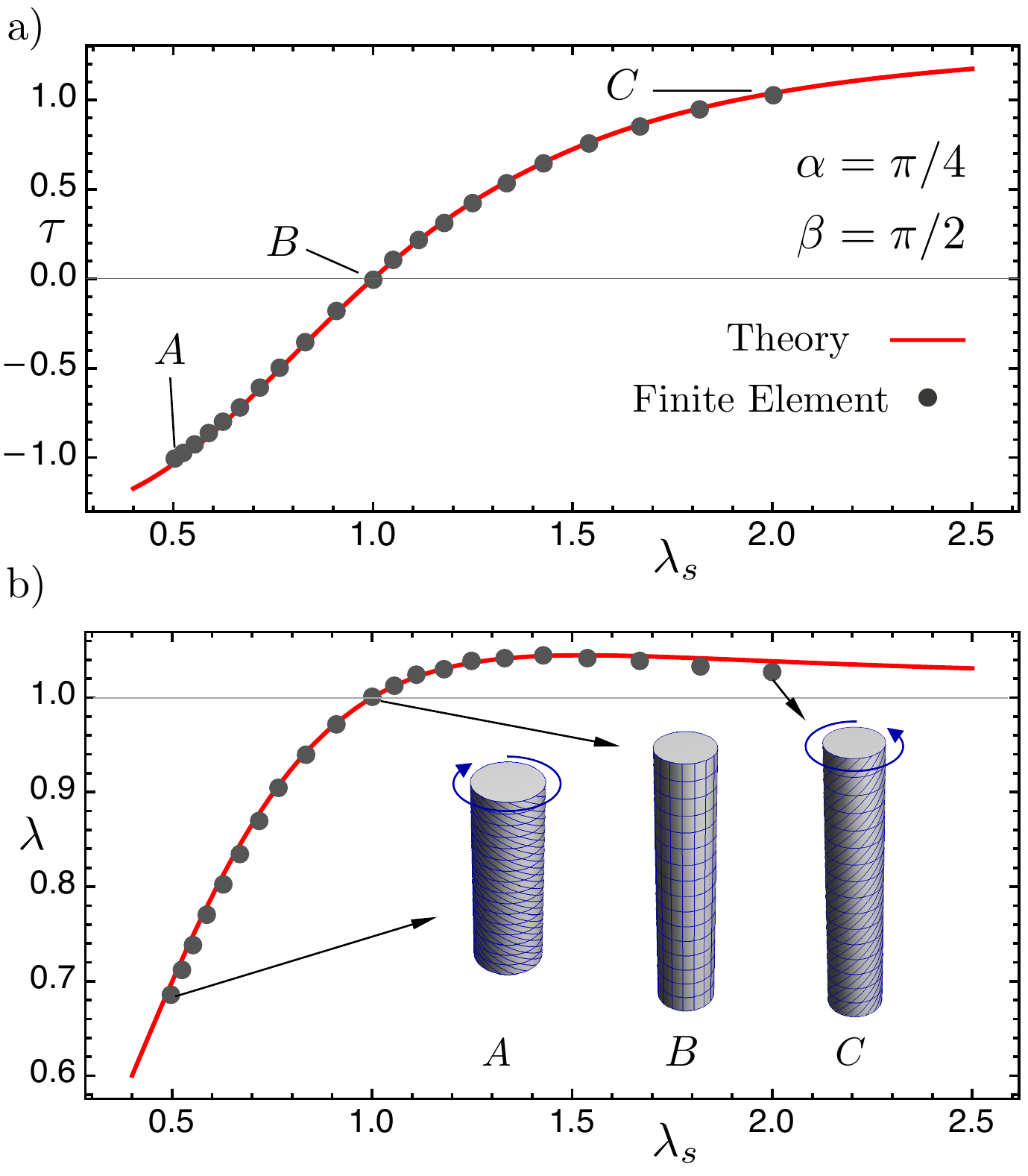}
\caption{\label{fig:stFE} Comparison between theoretical predictions and finite element simulation for twist (a) and stretch (b) as a function of the spontaneous deformation parameter $\lambda_s$. For these calcuations, we used a fibre with aspect ratio $R_0/L=1/15$ represented by $5400$ hex-8 elements. }
\end{figure}

\subsubsection{Finite element verification}
To test our results, we use the open-source finite element software \textit{FEBio} \cite{maas2012febio,maas2016general,maas2018plugin} to compute the spontaneous deformations of an LCE fibre encoded with an $R$-independent director field. We used FEBio's standard prestrained neo-Hookean material on cylindrical fibres with an almost incompressible Poisson ratio of $\nu= 0.45$. The isochoric prestrain parameter $1/\lambda_s$ was set to vary between values of $0.5$ to $2$. After applying the pre-strain, the energy minimising deformation was found using a static analysis. 

\begin{figure}[t]
\includegraphics[width=\columnwidth]{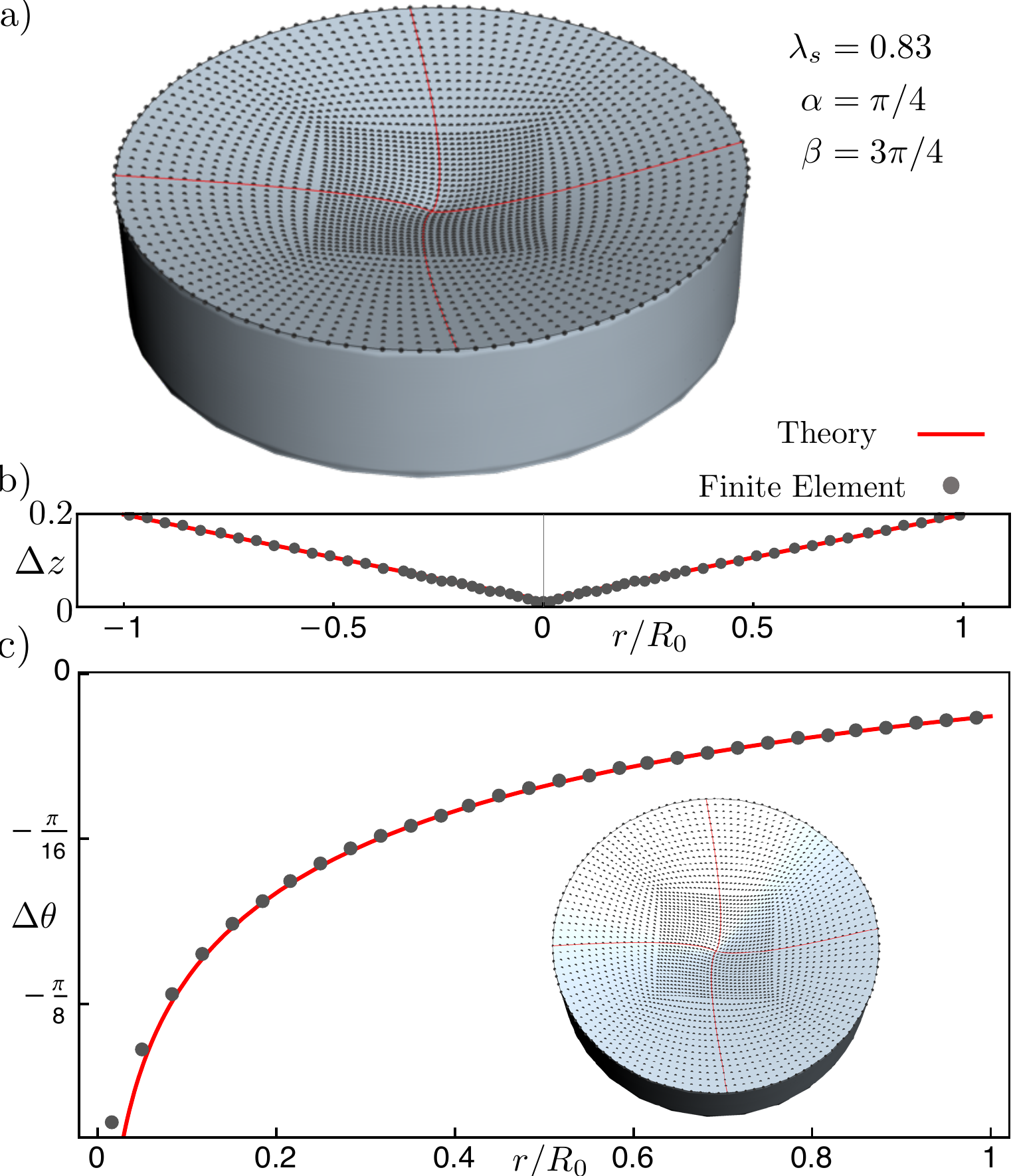}
\caption{\label{fig:coning} Comparison between Finite element simulations and theoretical data for the case $\alpha=\pi/4$ and $\beta=\pi/4$. a) An example of a deformed cross section directly from FE analysis. In red are theoretical lines for the deformed radii while dots are node positions. b) Comparison between predicted and theoretical coning. c) Comparison between theory and FE on the predicted rotation angle difference as a function of radius. A fibre with aspect ratio $R_0/L=1/5$, and $54000$ hex-8 elements was used for this simulation.}
\end{figure}

In figure \ref{fig:stFE}, we compare the predicted spontaneous twist and stretch of a fibre encoded with $\beta=\pi/2$ and $\alpha=\pi/4$ with finite element simulations. The theory shows excellent agreement with the numerical simulations, accurately capturing the non-linearities in both twist and stretch.

Fibres with $\beta=\pi/2$ have cross-sections that remain flat during deformation, as coning is driven by the $g_{R Z}$ component of the spontaneous deformation, which is only present if the director has an $RZ$ component. Therefore, to confirm our predictions about cross-section warping, we also computed the deformation of a fibre with $\beta=\pi/4$ and $\alpha=\pi/4$, as shown in Fig. \ref{fig:coning} a). Again, comparing our theoretical results with FE simulations we obtain excellent agreement between the numerical and theoretical coning (Fig. \ref{fig:coning} (b)) and winding (Fig. \ref{fig:coning} (c)) of the cross-section. The logarithmic nature of the spirals implies a theoretically infinite number of rotations at the centre of the cross section, although the stress, strain and energy are all finite. The divergent rotation stems from the line of director discontinuity (disinclination) along the central axis of the fibre. Accordingly, in a real fibre rotation would be cut-off near the axis by a regularisation of the director discontinuity within a defect core \cite{de1993physics}, and in our finite elements it is  cut-off by the element size near the axis. However, the director discontinuity line and associated infinite rotation are an artefact of patterns of constant $\alpha$ and $\beta$, and, as discussed in the next section, are not expected in the experimentally generated fibres.

\subsection{Nematic fibres produced by stretching and twisting during cross-linking}
\begin{figure}[t]
\includegraphics[width=\columnwidth]{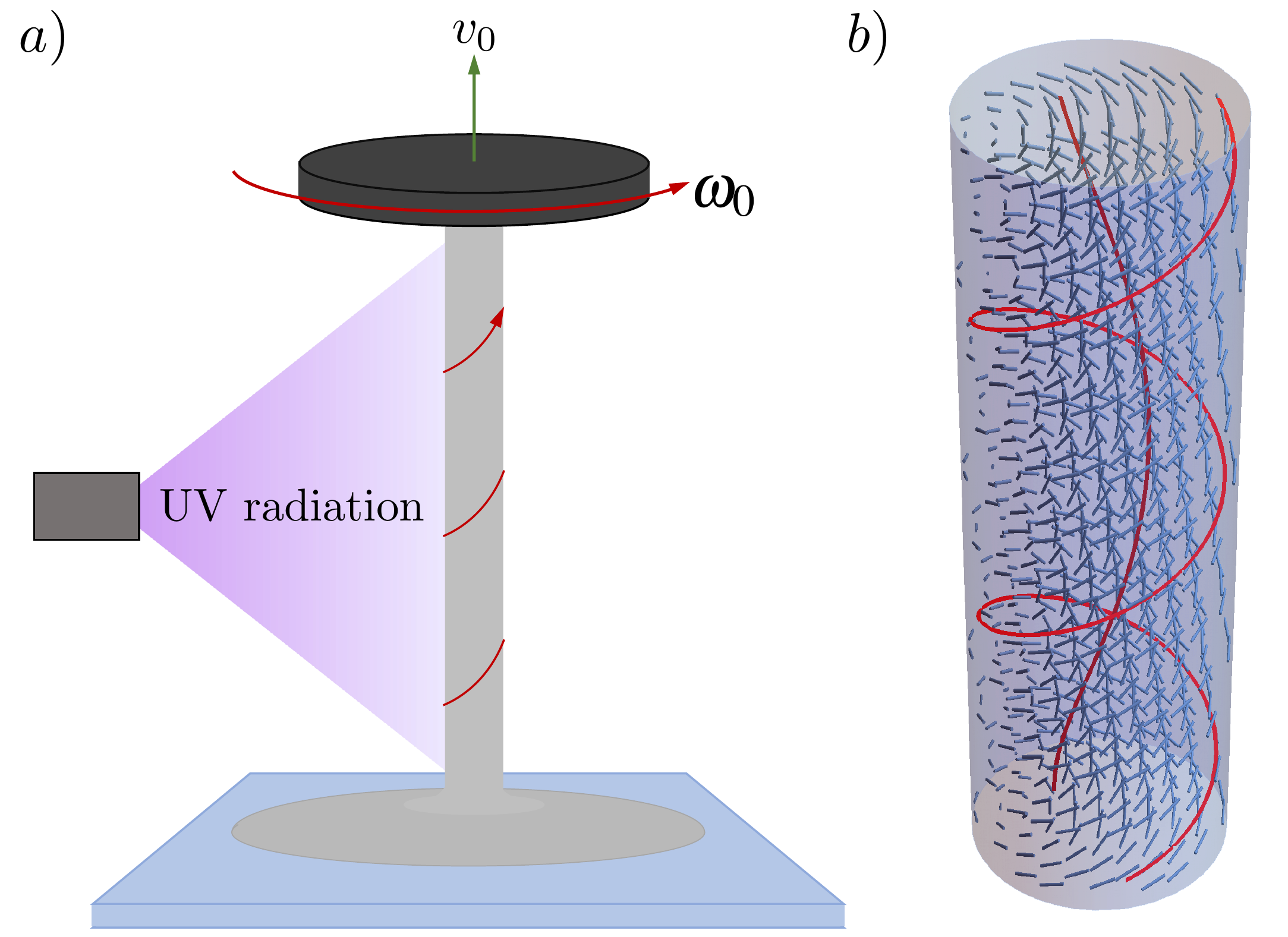}
\caption{\label{fig:gen2} a) Schematic of a fibre being drawn and twisted from  a drop of LC monomer while being cured with UV light. b) An example of the resulting director field in a fibre of radius $r=2/\tau_0$. The two integral curves highlight the change in azimuthal component as a function of the radius. }
\end{figure}

Finally, we consider the twisting LCE fibres reported by \emph{Nocentini et al}\cite{Nocentini2017b}. These fibres were produced by pulling a filament out of a viscous LC monomer mixture while rotating the drawing end, and simultaneously cross-linking with a UV light.  The director alignment is imprinted through the strains induced during this drawing process, shown in Figure \ref{fig:gen2} a). Given there is both twisting and stretching during crosslinking, and twisting strains are larger at larger radii, we  expect this fabrication to produce an $R$-dependent director field with azimuthal and longitudinal components.

In reality, the imprinting of the director-field is a complex visco-elastic process involving sticky polymers being cross-linked into a rubber. However, here, we take a simple approximation, and assume the deformation is mainly elastic, and the director aligns with the direction of maximum strain. The elastic approximation is clearly appropriate once there is sufficient crosslinking, but is probably also applicable to the initial visco-elastic drawing as the strain rate is rather high. 

To find the imprinted director pattern, we model the fibre during drawing as an elastic cylinder that is stretched by a factor of $\lambda_0$ and twisted by $\Delta \theta_0$ as cross-linking proceeds, resulting in a fibre of length $L$ and radius $R_0$ and (final state) twist density $\tau_0=\Delta \theta_0/L$. We use $(\tilde{R},\tilde{\Theta},\tilde{Z})$ as the reference state coordinates for this problem, so that we may use  $(R,\Theta,Z)$ for the final state coordinates, which then become the reference state coordinates in our spontaneous deformation analysis. The elastic deformation follows the simple treatment in  section \ref{sec:simpletwist}, leading, in our coordinate system, to the deformation fields
\begin{align}
\notag
R(\tilde{R},\tilde{\Theta},\tilde{Z})&=\tilde{R}/\sqrt{\lambda_0}\\
\notag
\Theta(\tilde{R},\tilde{\Theta},\tilde{Z})&=\tilde{\Theta}+\tau_0 \lambda_0 \tilde{Z}\\
\notag
Z(\tilde{R},\tilde{\Theta},\tilde{Z})&=\lambda_0  \tilde{Z}\\
\end{align}

To obtain the direction of maximum strain in the final (post cross-linking) configuration, we use the left-Cauchy deformation tensor
\begin{equation}
\tensor{b}=\tensor{F} \cdot \tensor{F} ^{T}=\left(
\begin{array}{ccc}
 \frac{1}{\lambda _0} & 0 & 0 \\
 0 & \frac{1}{\lambda _0}+\lambda _0^2 R^2 \tau _0^2 & \lambda _0^2 R \tau _0 \\
 0 & \lambda _0^2 R \tau _0 & \lambda _0^2 \\
\end{array}
\right).
\end{equation}
The largest eigenvalue of $\tensor{b}$ identifies the largest component of stretch while its corresponding eigenvector (which is a target state object) is its direction. Since this is the direction along which the director will orient, it can be used to express the values of the angles $\alpha$ and $\beta$ in the fibre. 
We trivially obtain that $\beta=\pi/2$, since the twisting during the manufacturing of the fibre induces no coupling of the $R$-$Z$ components. For $\alpha$, one obtains
\begin{equation}
\tan\alpha\!=\! \frac{ \left(\bar{\tau}_0^2-1\right)\!+\!\sqrt{ \left(\bar{\tau}_0^2+1\right){}^2+2 \lambda _0^{-3} \left(\bar{\tau}_0^2-1\right)+\lambda _0^{-3} }+\lambda _0^{-3} }{2 \bar{\tau}_0}
\end{equation}
where $\bar{\tau}_0=R \tau_0$. 
We note that, when the fibre is drawn from a drop, $\lambda_0 \gg 1$, which simplifies the angle to:
\begin{equation}
\alpha=\tan ^{-1}\left(R \,\tau_0\right).
\end{equation}
This implies that the director points along the $Z$ direction in the centre of the fibre and tilts in the $\Theta$-$Z$ plane as one moves outwards. This is a reflection of the fact that, during formation, $\Theta$-$Z$ shears grow like $R$ as the filament is drawn and twisted, thus inducing a greater azimuthal component further from the centre as shown in figure \ref{fig:gen2} b).

\subsection{Comparison between twist during cross-linking and twist during activation}
\begin{figure}[h]
\includegraphics[width=\columnwidth]{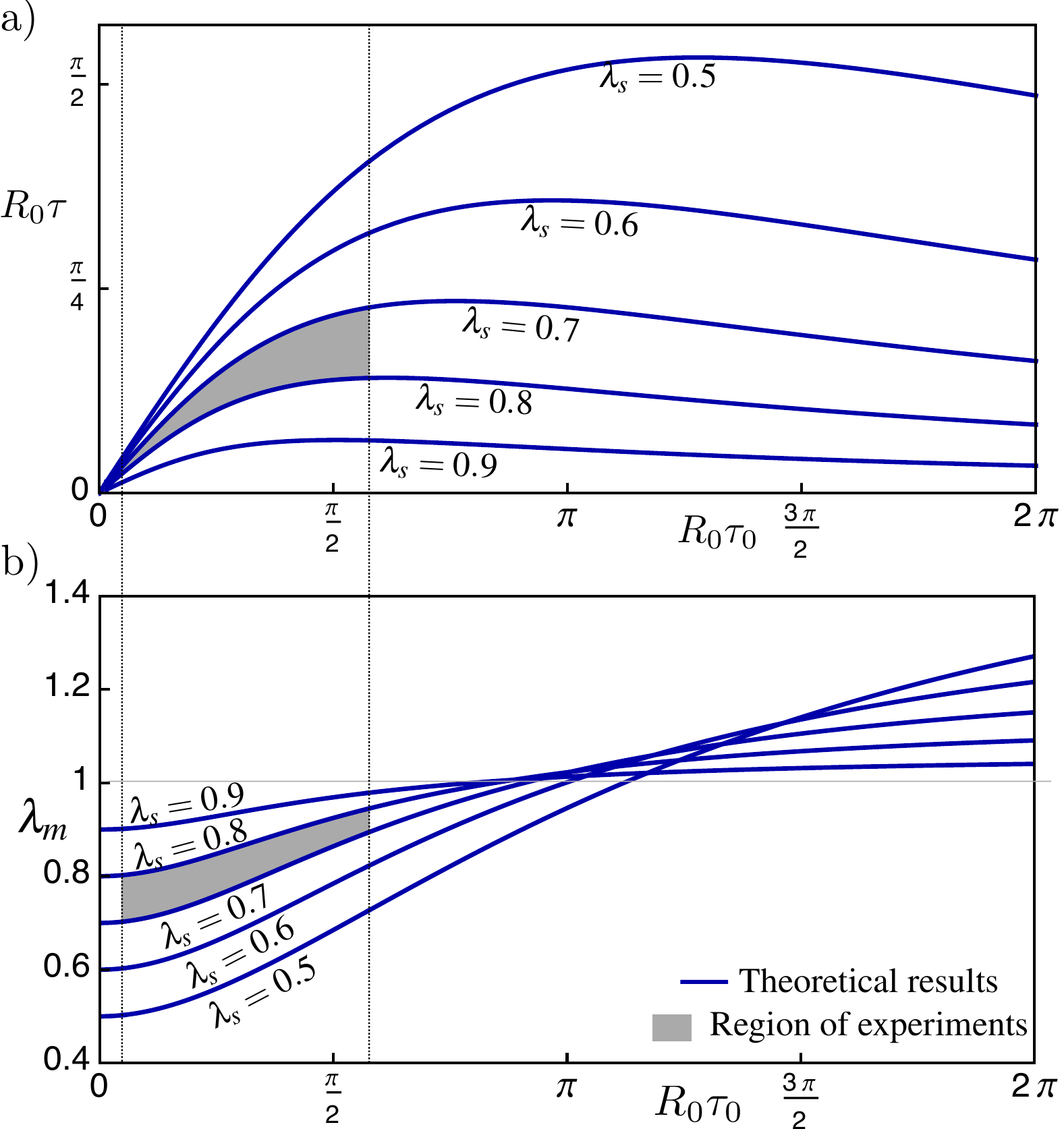}
\caption{\label{fig:twistcomp} Relationship between output twist a) and stretch b) as a function of the twist imposed at genesis. Shaded in grey, the region for known experimental results \cite{Nocentini2017b}.}
\end{figure}

Finally, we can use the form of $\alpha$ and $\beta$ to obtain $\tensor{g}$. We then use equations \eqref{coeff1} and \eqref{minimisers} to find the twist and stretch capability of a fibre given the twist imposed at its genesis. The results for values of $\lambda_s<1$ are shown in figure \ref{fig:twistcomp}. 
Remarkably, the output twist does not monotonically grow as a function of $\tau_0$, but reaches a maximum and decays to zero thereafter. 
Recall that, in the $R$-independent field, we discussed how the twist is maximised when $\alpha\sim \pi/4$. In this case, a small $\tau_0 R_0$ implies the director is on average mainly longitudinally aligned ($<\alpha> \sim 0$), inducing mostly a contraction by a factor $\lambda_s$. On the other hand, a large $\tau_0 R_0$ leads to a dominantly azimuthal director orientation on the cross-section ($<\alpha> \sim \pi/2$), inducing mainly stretch by a factor of $\left(\half(\lambda _s^3+1)/ \lambda _s^3\right)^{1/3}$. The optimum twist output $\tau$ is maximised in between the two, when the coupling between azimuthal and longitudinal component is greatest. 

The fibres produced by \emph{Nocentini et al}\cite{Nocentini2017b} were made from LCEs capable of a maximum spontaneous contraction of $\lambda_s\sim 0.71$ during heating. The twist imposed at genesis on the fibres was of about $10$ turns with their diameter and length varying between $50 $ and $300 \mathrm{\mu m}$ and $1$ to $5 \mathrm{cm}$ respectively.  This suggests their fibres were fabricated with a (dimensionless) genesis twist of around $0.16<R_0\tau_0<1.8$, and the corresponding experimentally-explored region is shaded on Fig.\ \ref{fig:twistcomp}. The authors only reported the output twist and contraction for one fibre (of unknown length and radius), when activated in a $\sim 0.5 \mathrm{cm}$ long region via light. This fibre generated an overall contraction $\sim 0.84$, suggesting that in the activated region $0.71<\lambda_m<0.84$. On the other hand, activation induced a rotation of about $\Delta \theta\sim 460^o\sim 8.2\,\mathrm{rad}$ corresponding to an output twist of  $0.04<R_0 \tau< 0.5$. Both these twist and stretch values fall in the shaded region of Fig.\ \ref{fig:twistcomp}, consistent with our theoretical results. We highlight how, in general, these experimental fibres appear to have been generated with too little twist at genesis, yielding a sub-optimal output twist. This could perhaps be improved by increasing the number of turns during fabrication. 

\section{conclusion and discussion}
In conclusion, we have obtained an exact elastic solution for a twisted and stretched cylindrical neo-Hookean fibre subject to an axisymmetric isochoric spontaneous deformation field. The solution yields a full non-linear elastic energy for such a fibre, which is minimized by a spontaneous twist and stretch. The energy also highlights a large-deformation coupling between twist and stretch, as familiar from the classical Poynting effect. Finally, the elastic fields also capture the large deformations of the fibre's cross-section, which is predicted to warp into a surface of revolution and twist such that radii become curves during activation.

When applied to LCEs, our results show that a helical director-field oblique in the longitudinal-azimuthal plane is required to induce twist. The twist output depends on the spontaneous elongation $\lambda_s$ and is maximised when no radial director component is present ($\beta=\pi/2$) as well as when the azimuthal and longitudinal components are coupled through an angle $\alpha\sim\pi/4+\threequarters (\lambda_s-1)$.

It is instructive to compare our theory with recent work on the spontaneous bending and twisting of rods via incompatible (aka misfit) spontaneous distortions\cite{Aharoni2012,Cicalese2017,Bauer2020,Kohn2018a,Moulton2020}. These treatments go beyond ours in that the spontaneous strain is not assumed to be axisymetric, and the rod is allowed to bend so that the center line no-longer remains straight. However, these treatments do assume small spontaneous strains and high aspect-ratio rods, allowing a linear elastic treatment similar to the original Kirchhoff model. The small-strain thin-rod regime limits the theories to stretch free deformations, and leads to simple bend twist energies of the form
\begin{equation}
   \frac{\mathcal{E}}{\mu \pi L R_0^2}= R_0^2  \left(\threeeights(\vec{\kappa}-\vec{\kappa}_m)^2+\fourth(\tau-\tau_m)^2\right)\label{eq:linear}
\end{equation}
where $\vec{\kappa}$ and $\tau$ are the bend (curvature vector) and twist of the rod, while $\vec{\kappa}_m$ and $\tau_m$ are their minimising values. These minimising values were first estimated by linearising the spontaneous deformation in a Taylor series about the rod's central axis \cite{Aharoni2012}. More recently, Gamma convergence\cite{Cicalese2017,Bauer2020,Kohn2018a} and 3D energy minimization\cite{Moulton2020} have been used to derive rigorous forms, yielding averages and moments of various terms of the spontaneous deformation over the rod's cross-section. Our large-deformation axisymmetric treatment reduces to the twisting portion of these  rod-theories in the limit of small spontaneous distortions and little imposed stretch. Indeed, if we Taylor expand our expressions for $\lambda_m$ and $\tau_m$ (eqn.\ \eqref{minimisers}) in the limit of small spontaneous deformations $\tensor{g}=\tensor{\delta}+\epsilon \tensor{\delta g}(R)$, we find that 
\begin{align}
    \lambda_m&=1+ \epsilon \frac{1}{R_0^2}\int_0^{R_0}  \left(\delta\!g_{RR}+\delta\!g_{\Theta \Theta}\right) R\, dR\\
 \tau_m&= -\epsilon\frac{4 }{R_0^4}\int_0^{R_0} \delta\!g_{\Theta Z} \, R^2dR.
\end{align}
If we then also assume the applied twist and stretch are small, $\lambda-\lambda_m \sim \epsilon$ , $\tau\sim \epsilon$, we may expand our energy (Eqn.\ \eqref{eqn:energymain}) to $\epsilon^2$ to obtain
\begin{equation}
    \frac{\mathcal{E}}{\mu \pi L R_0^2 }=\threehalf \left( \lambda -\lambda _m\right)^2+\fourth R_0^2 \left(\tau -\tau _m\right)^2+\mathcal{O}(\epsilon^3).
\end{equation}
The twisting portion of this energy agrees with that in eqn.\ \eqref{eq:linear}, and the linearized form of $\tau_m$ above agrees with that  given by \emph{Kohn and O’Brien}\cite{Kohn2018a}. Interestingly, although such small-strain and high-aspect ratio assumptions appear necessary to resolve bending, our treatment demonstrates they can be avoided entirely when only treating twisting and stretching. The resulting non-linear formulation generates the highly non-linear form $\tau_m(\lambda_s)$ and $\lambda_m(\lambda_s)$, as seen in fig.\ \ref{fig:stFE}, and is clearly required for accurate predictions in large strain systems such as LCEs. 

Remarkably, even a small strain rod theory can describe large bend and twist displacements in a suitably long rod. This introduces fascinating and rich geometric coupling between twist and bend deformations \cite{VanDerHeijden2000,Charles2019,Fuller1971,Purohit2008}. For example, if one straightens a wound headphone wire, it becomes highly twisted. Similarly, if one twists a fibre then brings the ends together, it will spontaneously untwist into a lower energy spring-like coiled state.  In general, torsional bend and twist can be exchanged in a rod or fibre, without rotating the ends, provided the total number of turns is conserved. This  twist-torsion coupling is an example of a geometric phase and is key in the winding of DNA molecules \cite{Marko1998,Olsen2011,Fuller1978}, and the coiling of plant tendrils\cite{Gerbode2012}.

\begin{figure}[t]
\includegraphics[width=\columnwidth]{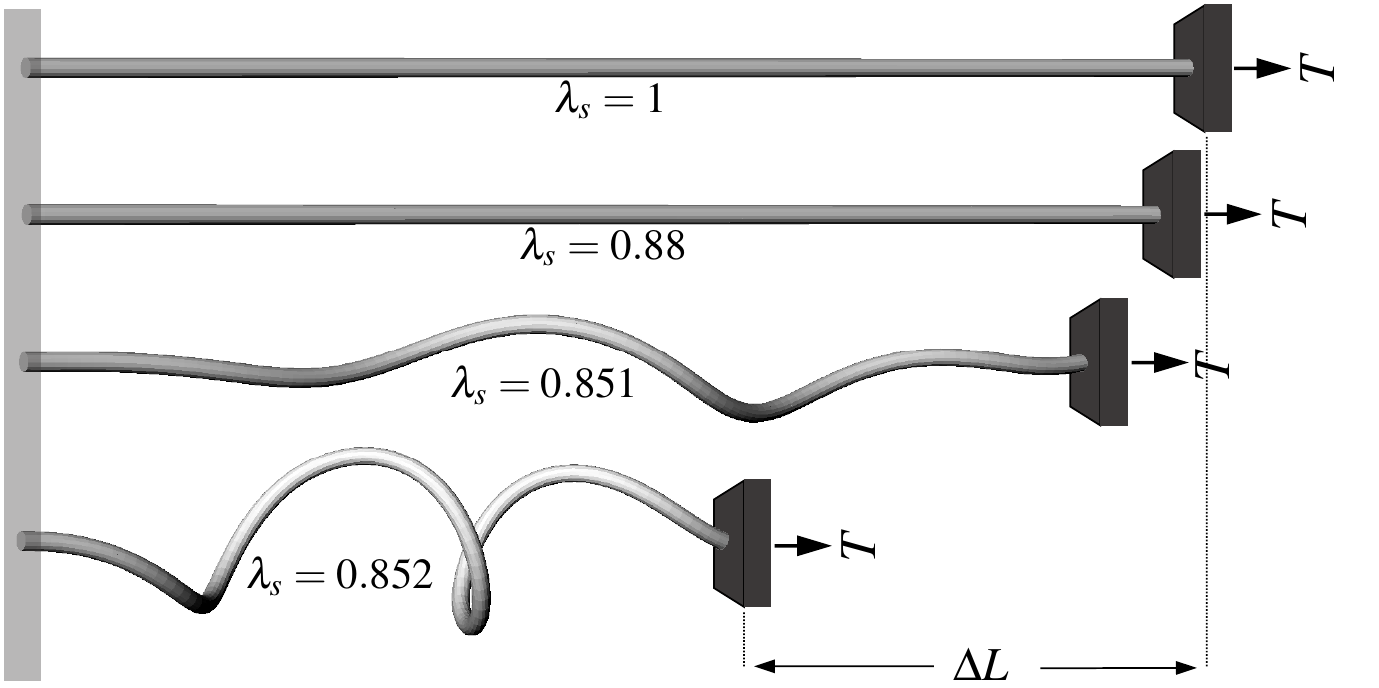}
\caption{\label{fig:coil} Finite element simulation of a fibre with Young modulus $E=1$, Poisson ratio $\nu=0.45$, 4800 elements, aspect ratio of $R_0/L=0.07$ and a $R$ independent director-field with $\alpha=\pi/4$ and $\beta=\pi/2$. 
The fibre is fixed at one end and subject to a tension $T=0.2 \pi \mu R_0^2$ while preventing rotation at the other end. As the value of $\lambda_s$ is decreased, the fibre wants to shrink and twist but cannot do the latter as the ends are not allowed to rotate. The fibre thus coils to  release some of the twist energy, resulting in a greater stroke amplitude then that of classical linear actuation. }
\end{figure}

Therefore, although our treatment includes no mention of bend, we can infer from this coupling effect that a spontaneously twisting LCE fibre will bend into a coiled spring-like configuration if it activates under boundary conditions that prevent the ends from twisting relative to each other. Given the large-strain nature of LCE actuation, a formal treatment of this torsional effect appears to be a formidable challenge. However, as shown in Fig.\ \ref{fig:coil}, we were easily able to observe it in finite elements simulations of a twisting LCE fibre actuated under a constant longitudinal force, but with the constraint that the ends may not rotate. This coiling mechanics is commonly deployed in other twisting artificial muscles to generate linear actuation \cite{aziz2020torsional,Charles2019}. In contrast, LCEs are intrinsically contractile actuators, and simple linear contraction can be trivially achieved with a monodomain strip \cite{kupfer1991nematic,de1997artificial,wermter2001liquid}. However, the coiling mechanism would allow an LCE actuator with much higher stroke and lower stiffness, forming switchable soft-springs that, like plant tendrils\cite{Gerbode2012}, could be used to gently anchor and position an object in 3-D space.

\acknowledgements{A.G.\ thanks J.L. for the support and the EPSRC for funding, project 2108804. J.S.B.\ is supported by a UKRI Future Leaders Fellowship MR/S017186/1.}
\\\\
 \textbf{DATA AVAILABILITY}
 The data that supports the findings of this study are available within the article.
 \appendix
  \section{Modelling a nematic elastomer as a neo-Hookean with spontaneous distortion \label{appendixa}}
The free energy density of a nematic elastomer \cite{warner2007liquid} has its roots in statistical physics, and contains two contributions: an elastomer energy from the polymer network, and a nematic energy from the rods, 
\[
W=W_{pol}+W_{rod}.
\]
The rod energy depends on the scalar order parameter $Q$ of the nematic field (i.e. the degree of alignment) but not its direction (unit vector $\vec{n}$). Appropriate forms for $W_{rod}(Q)$ are provided by the Landau-de-Gennes  (phenomenological) or Maier Saupe (microscopic) theories of liquid nematics \cite{de1993physics}. Either way,  $W_{rod}(Q)$ has characteristic size of $k_B T$ per rod, and transitions from having a minimum at $Q=0$ (isotropic) to a finite $Q$ (nematic) below a critical temperature $T^*$.

The polymer free energy is dominated by conformational entropy, like in the statistical theory of  conventional rubber. However, in the presence of a nematic field, the polymer random walks are not isotropic, but biased along the director $\vec{n}$ by an amount $r(Q)$ determined by the degree of alignment, as encoded in a a step-length tensor $\tensor{\ell} \propto \delta+(r-1)\vec{n}\vec{n}$. The resultant polymer energy is described by the ``trace formula'' \cite{bladon1993transitions, warner2007liquid} 
\[
W_{pol}=\half n_s k_B T \,\Tr{\tensor{\ell}_0 \cdot \tensor{F}^{T} \cdot \tensor{\ell}^{-1} \cdot \tensor{F}}
\]
where $n_s$ is the density of polymer strands, $\tensor{F}$ is the deformation from the cross-linking state to the final state, $\ell_0$ is the step length tensor at cross-linking (which depends on the nematic variables at cross-linking, $Q_0$ $\vec{n_0}$) and $\tensor{\ell}$ is the step-length tensor in the final state (which depends on the final state nematic variables at $Q$, $\vec{n}$). The full behaviour of the nematic elastomer is now given by minimizing the sum of both energies over elastic deformations ($\tensor{F}$), final state order parameter ($Q$), and final state director ($\vec{n}$). In general, this minimisation gives a two way coupling between the nematic order and the LCE deformation. However, the characteristic size of $W_{rod}$ is $k_B T$ per rod, while the characteristic size of $W_{pol}$ is $k_B T$ per polymer strand. Since in an elastomer there are typically more than ten rods per strand, the nematic energy dominates the elastic one during minimization over $Q$. Therefore, the degree of nematic alignment is essentially that which minimizes the nematic energy alone, and is only modestly affected by the polymer network. For example, the shift of the nematic-isotropic transition caused by the presence of the network is typically only a few Kelvin \cite{zubarev1997phase} (compared to a transition temperature of ~350K), and other mechanically induced changes in $Q$ are similarly small \cite{warner2007liquid,kaufhold1991nematic}. Such effects are more pronounced in some modern LCE compositions which have more crosslinks per rod \cite{jampani2019liquid, mistry2018coincident}, but, for simplicity, we focus on the traditional case here.

 We may thus consider the simpler problem of minimizing the polymer energy, with the magnitude of nematic order, $Q$, effectively fixed as a constraint by the rods energy. In general, when we minimize the polymer energy in the nematic state, we must still do it over final state director $\vec{n}$; indeed the nematic director can be observed to rotate within an elastomer in response to stretch \cite{warner2007liquid, finkelmann1997critical}. However, if we further assume (as here) that the final state is isotropic, $Q=0$, then $\ell\propto \tensor{\delta}$ must also be isotropic, and there is no final state director to minimize over. In this case, the elastomer energy is simply
  \[
W=\half n_s k_B T \,\Tr{\tensor{F} \cdot \tensor{\ell}_0 \cdot \tensor{F}^{T} }
\]
which corresponds to the standard neo-Hookean energy with spontaneous distortion $\tensor{G}=\ell_0^{1/2}$ encoded by the nematic field in the cross-linking state and shear modulus $\mu=n_s k_B T$. 

\bibliography{TM}

\end{document}